\documentclass[journal]{IEEEtran}

\usepackage[T1]{fontenc}
\usepackage[utf8]{inputenc}
\usepackage{amsmath,amssymb}
\usepackage{booktabs}
\usepackage{graphicx}
\usepackage{xcolor}
\usepackage{array}
\usepackage{tabularx}
\usepackage{multirow}
\usepackage{siunitx}
\usepackage{cite}
\usepackage{url}
\usepackage[nospread]{flushend}
\usepackage[hidelinks]{hyperref}
\hypersetup{
  pdftitle={Feeder-Aware Service-Conditioned Flexibility of Grid-Interactive Buildings},
  pdfauthor={Dafang Zhao, Ittetsu Taniguchi, Takao Onoye}
}

\graphicspath{{figures/}}
\definecolor{pilotblue}{RGB}{0,92,138}
\newcommand{\Fd}{F^{\mathrm{demand}}}
\newcommand{\Fg}{F^{\mathrm{grid}}}
\newcommand{\Fcap}{F^{\mathrm{grid,cap}}}
\newcommand{\Fdlo}{F^{\mathrm{demand},-}}
\newcommand{\Fdhi}{F^{\mathrm{demand},+}}
\newcommand{\Fglo}{F^{\mathrm{grid},-}}
\newcommand{\Fghi}{F^{\mathrm{grid},+}}

\title{Feeder-Aware Service-Conditioned Flexibility of Grid-Interactive Buildings}

\author{Dafang Zhao,~\IEEEmembership{Member,~IEEE,}
        Ittetsu Taniguchi,~\IEEEmembership{Senior~Member,~IEEE,}
        and Takao Onoye,~\IEEEmembership{Senior~Member,~IEEE}%
\thanks{Corresponding author: Dafang Zhao, zhao.dafang@ist.osaka-u.ac.jp}}

\begin{document}

\maketitle

\begin{abstract}
Grid-interactive buildings can adjust demand, but building-feasible power budgets may differ from accepted feeder-head offers.
This study addresses a practical question: to what extent can the adjustable electricity consumption of buildings be translated into deliverable flexibility services while satisfying feeder-head operating requirements? 
To enable a fair comparison, we adopt the same baseline operating schedule, equipment conditions, and post-service recovery requirements for indoor temperature and battery state of charge (SOC). 
Flexibility is then evaluated under two settings: one considering only building-level constraints, and the other additionally accounting for feeder-head constraints.
We first use a simplified network model with explicit security margins to determine a preliminary amount of flexibility that can be committed. 
The resulting schedule is then validated using a more detailed network simulation via Modelica. 
If network losses cause the actual response to deviate from the committed response, the electricity consumption of individual buildings is rescheduled.
If the required response still cannot be achieved, the committed flexibility level is reduced.
The experiments cover multiple service activation times, regulation directions, and service durations. 
For comparative purposes, we retain the initial optimization result as the reference estimate of the maximum potentially available flexibility, rather than interpreting the failure of a particular subsequent realization attempt as evidence that a larger flexibility level is inherently infeasible. 
The results demonstrated that 24.51\% of nonzero flexibility contracts were rejected due to feeder-head constraints, 67.04\% of contracts were accepted with some rescheduling and building redispatch, and only 8.09\% of contracts were accepted without any changes.
\end{abstract}


\section*{Nomenclature}
\begingroup
\small\raggedright
\renewcommand{\IEEEiedlistdecl}{%
  \IEEEsetlabelwidth{$\Fdlo,\,\Fdhi$}%
  \IEEEusemathlabelsep\IEEElabelindent=0pt\relax}
\begin{IEEEdescription}
  \item[$b,\,\mathcal{B}$] Building index and population.
  \item[$i,j,\,\mathcal{L}$] Bus indices and oriented branch set.
  \item[$k,K,\Delta t$] Interval index, terminal index, and time step (0.25 h).
  \item[$\tau,D,s$] Activation time, duration, and direction ($+1$ upward, $-1$ downward).
  \item[$T,E,p$] Zone temperature, stored battery energy, and net import; superscript $0$ denotes reference operation.
  \item[$R_b,C_b$] Thermal resistance (K/kW) and capacitance (kWh/K).
  \item[$P_{ij,k},Q_{ij,k},v_{j,k}$] Per-unit branch powers and squared bus voltage.
  \item[$P_{0,k},P^{\mathrm{AC}}_{0,k}$] Lossless aggregate and nonlinear source import (kW).
  \item[$\ell_k,r^{\mathrm{AC}}_k,\delta_k$] Feeder loss, metered directional response, and aggregate setpoint offset (kW).
  \item[$F,A_0$] Service amplitude and initial feasible linear proposal (kW).
  \item[$\Fd,\Fg_{\mathrm{lin}}$] Optima of the paired lossless building and linear feeder problems.
  \item[$\Fcap$] Maximum source amplitude subject to the proposal budget $A_0$, linear restrictions, and nonlinear acceptance.
  \item[$\Fdlo,\Fdhi$] Feasible lower endpoint and upper bound on the lossless building capacity.
  \item[$\Fglo,\Fghi$] Accepted lower endpoint and retained primary ceiling bounding $\Fcap$.
  \item[$\rho^-,\rho^+,e$] Bounds on conditional retention $\Fcap/\Fd$ and corresponding erosion.
  \item[$\mathcal{B}_c,\mathcal{S}^0(F)$] Resource/recovery set and zero-offset lossless service equality.
  \item[$\mathcal{L}_c,\mathcal{N}(F)$] Linear feeder restrictions and nonlinear source-acceptance tests.
\end{IEEEdescription}
\endgroup

\section{Introduction}
\IEEEPARstart{P}{ower-system} decarbonization increases the value of demand that can adapt to variable renewable generation and changing peak loads \cite{oshaughnessy2022demand}.
Buildings can coordinate heating, ventilation, and air-conditioning (HVAC) loads, batteries, and photovoltaics (PV) as distributed energy resources (DERs) for demand response \cite{li2021review}.
Aggregation enables small resources to satisfy service requirements collectively.
For example, the U.S. Federal Energy Regulatory Commission's Order No.~2222 established a route for aggregated DER participation in regional wholesale markets.
Its locational and distribution-utility coordination provisions highlight the need to determine how much pooled flexibility the local network can deliver.

Aggregation faces coupled temporal and spatial constraints.
At the device level, instantaneous power availability does not establish a sustainable response.
Batteries can exhaust stored energy, while thermal loads approach comfort limits \cite{stinner2016thermal,bampoulas2021framework}.
Recovery also consumes flexibility; recharging and thermal rebound can shift demand beyond the activation period \cite{kathirgamanathan2020indicators}.
An offer therefore requires a feasible sequence of device actions throughout service and recovery, rather than a sum of independently available power limits.

At the network level, different building allocations can produce the same aggregate power change but different branch flows and voltage profiles \cite{fontenot2021b2dn,navidi2023reliability}.
Individually feasible responses may jointly overload shared conductors or transformers.
Upward service reduces import or increases export, potentially relieving import loading or worsening reverse flow.
Downward service increases import or absorbs generation, which can relieve PV-driven overvoltage but worsen undervoltage during high demand.
Thus, flexibility depends on location, operating conditions, direction, and duration.
Ignoring these dependencies can cause an aggregator to overstate its offer or shift stress from activation to recovery.

Research has addressed these challenges through complementary approaches.
Building-flexibility metrics and resource models characterize reference-dependent power, energy, and duration limits \cite{li2023kpi,tang2021flexibility}.
Trajectory-independent envelopes protect load feasibility across admissible trajectories \cite{rousseau2026envelope}.
Integrated building--network control coordinates dispatch against electrical constraints \cite{mirakhorli2018mpc,jiang2018adn,fontenot2021b2dn,zhang2022transformer}.
Network operating envelopes allocate admissible exchanges \cite{liu2021operating,gerdroodbari2022pq,liu2022opf}.
Recent combinations of these approaches already support comfort-aware demand response and deliverable building services \cite{lankeshwara2025doe,kheiri2026deliverable}.
Together, these advances establish methods for feasible coordination and service delivery.
What is less clearly quantified is: under exactly the same building conditions and recovery requirements, how much of the flexibility that appears available at the building level can actually be converted into a deliverable feeder-head service?

This study assesses the key question that how much of building side flexibility can truly be converted into regulation services deliverable at the feeder head while satisfying the operational constraints of grid.
To separate the ``Building side constraints'' and ``Feeder-head constraints'', this paper individually calculated the flexibility considering only building constraints, and the deliverable flexibility when further accounting for feeder constraints.
We conducted a long-term experiment include over 4,000 different startup time, dual direction flexibility adjustment and three different flexibility services durations.
Therefore, this study systematically examines the process of transforming building flexibility into services deliverable to grid across varying times, directions, and durations.
Within this framework, a natural question arises: regarding the transformation from ``building-side flexibility'' to ``feeder-head deliverability'' exactly how much capacity is lost, when does the loss occur, and why? Furthermore, what is the cost of the adjustments required to ultimately achieve deliverable services?
Therefore, this paper further proposes four research questions:
\noindent\textit{RQ1:} What fraction of the building-side budget is retained under the linear proposal budget and nonlinear acceptance conditions?
\noindent\textit{RQ2:} How does this conditional retention vary with activation time, duration, and seasonal subset?
\noindent\textit{RQ3:} Which physical limits reject demand-only trajectories, and which checks reject initial feeder candidates?
\noindent\textit{RQ4:} How much redispatch, amplitude reduction, and computation are needed to obtain accepted responses?
To address these questions, this paper not only provide the final flexibility value but also established a comprehensive assessment framework that capable of distinguishing building-side capabilities, impact of grid constraints, and non-linear delivery errors at the sources.
Based on this approach, the main contributions of this paper can be summarized in the following three aspects.
\begin{enumerate}
  \item A paired assessment with an explicitly bounded target that separates lossless capacity optimization from source-side acceptance under the resulting proposal budget.
  \item A six-month analysis of conditional retention and network exposure that distinguishes linear-budget restrictions from tracking-driven correction.
  \item An auditable reconstruction of proposal ceilings, accepted trajectories, loss compensation, and numerical recovery, with interval widths and correction mechanisms reported separately.
\end{enumerate}

\section{Related Work}

\begin{table*}[htbp]
    \caption{Selected related approaches and their relationship to the paired service-capacity assessment.}
    \label{tab:positioning}
    \centering
    \footnotesize
    \setlength{\tabcolsep}{4pt}
    \renewcommand{\arraystretch}{1.06}
    \begin{tabularx}{\textwidth}{@{}>{\raggedright\arraybackslash}p{34mm}>{\raggedright\arraybackslash}p{65mm}>{\raggedright\arraybackslash}X@{}}
    \toprule
    Study & Established capability & Boundary relevant to this comparison \\
    \midrule
    Kathirgamanathan et al.; Bampoulas et al. \cite{kathirgamanathan2020indicators,bampoulas2021framework} & Quantity, efficiency, and rebound indicators; flexibility maps for electrical and thermal resources. & Building response potential requires a separate check against shared feeder limits. \\
    Rousseau et al. \cite{rousseau2026envelope} & Trajectory-independent energy bounds for loads with state-dependent losses. & Load-feasibility guarantees and electrical-network deliverability are distinct assessments. \\
    Fontenot et al.; Zhang et al. \cite{fontenot2021b2dn,zhang2022transformer} & Integrated building-network control with comfort, voltage regulation, transformer loading, and reconfiguration. & Operational objectives differ from maximizing capacity under identical service and recovery conditions. \\
    Bay et al.; Yu et al. \cite{bay2022dmpc,yu2024online} & Distributed aggregate tracking or peak-demand coordination with local building objectives. & Aggregate coordination does not itself measure the fraction of capacity removed by feeder constraints. \\
    Liu et al.; Gerdroodbari et al. \cite{liu2022opf,gerdroodbari2022pq} & Network-admissible meter-level import/export or active-reactive power envelopes. & Sustained service also requires intertemporally feasible building and storage trajectories. \\
    Lankeshwara et al. \cite{lankeshwara2025doe} & Envelope-enabled demand-response tracking, comfort protection, and software-in-the-loop validation. & Validated target tracking motivates a separate comparison of matched capacity bounds. \\
    Kheiri et al. \cite{kheiri2026deliverable} & Hierarchical day-ahead coordination of building resources and deliverable grid services. & The present comparison instead fixes the reference and recovery conditions while adding feeder constraints. \\
    Langner et al.; Zhang et al. \cite{langner2024uncertainty,zhang2021stochastic} & Uncertainty-aware building control and stochastic network voltage regulation. & Forecast-risk treatment remains outside the present perfect-information capacity assessment. \\
    This study & Paired lossless budgets and accepted source offers across six months, two directions, and three durations. & Bounds concern the stated proposal budget and conservative formulation; no controller-superiority claim. \\
    \bottomrule
    \end{tabularx}
\end{table*}

Building-side methods characterize thermal and storage dynamics through power, energy, shifting efficiency, and rebound indicators \cite{stinner2016thermal,kathirgamanathan2020indicators}.
Integrated frameworks map flexibility across electrical and thermal resources.
Weather and occupancy shape resource availability \cite{bampoulas2021framework}.
Coordinating local electricity and heat supply provides additional consumer-side flexibility \cite{zhang2018consumer}.
Intertemporal feasibility also depends on how a flexibility envelope is constructed.
Rousseau et al. show that extremal cumulative-energy bounds can admit power trajectories that violate load constraints when losses depend on state \cite{rousseau2026envelope}.
Their trajectory-independent bounds address this problem for individual and physically coupled loads.
However, these guarantees concern internal load feasibility.
Deliverability through an electrical network requires an additional assessment of the resulting injections.

Joint building-network optimization integrates building flexibility into coordinated electrical feeder operation \cite{mirakhorli2018mpc}.
Fontenot et al. coordinate buildings, PV, and batteries with voltage regulation, reactive support, and comfort constraints \cite{fontenot2021b2dn}.
Their study also includes multiple feeder configurations and uncertainty analysis.
Related studies incorporate transformer loading and reconfiguration \cite{zhang2022transformer} or HVAC demand response for photovoltaic hosting capacity \cite{hwang2023hosting}.
Dynamic operating envelopes communicate time-varying network-admissible import or export limits while allowing customers to manage their resources \cite{liu2021operating,liu2022opf}.
Extensions include active-reactive power regions for PV and batteries \cite{gerdroodbari2022pq} and negotiated envelopes for decentralized trading with information privacy \cite{jiang2025trading}.
These methods allocate network capacity, but sustained service also requires feasible thermal and storage trajectories during activation and recovery.

Recent studies explicitly connect operating envelopes and building constraints to service delivery.
Lankeshwara et al. develop two-stage envelope allocation and hierarchical demand-response control, with software-in-the-loop validation of tracking, network limits, and thermal comfort \cite{lankeshwara2025doe}.
Kheiri et al. coordinate heat pumps, water heating, PV, and storage through local aggregators and a distribution-system operator \cite{kheiri2026deliverable}.
Their day-ahead framework combines occupant comfort with voltage, line-capacity, and loss considerations to provide deliverable grid services.
These studies establish that network-aware, comfort-constrained delivery is already achievable.
In the envelope-enabled formulation, connection-level active--reactive power regions constrain a controller tracking a supplied aggregate request \cite{lankeshwara2025doe}.
Here, the paired lossless problems treat the request amplitude as a decision variable.
A separately defined source-side acceptance search then operates within the resulting budget.
This changes the assessment objective and comparator without asserting a new control principle or superiority over those controllers.

\begin{figure*}[!t]
    \centering
    \includegraphics[width=\textwidth]{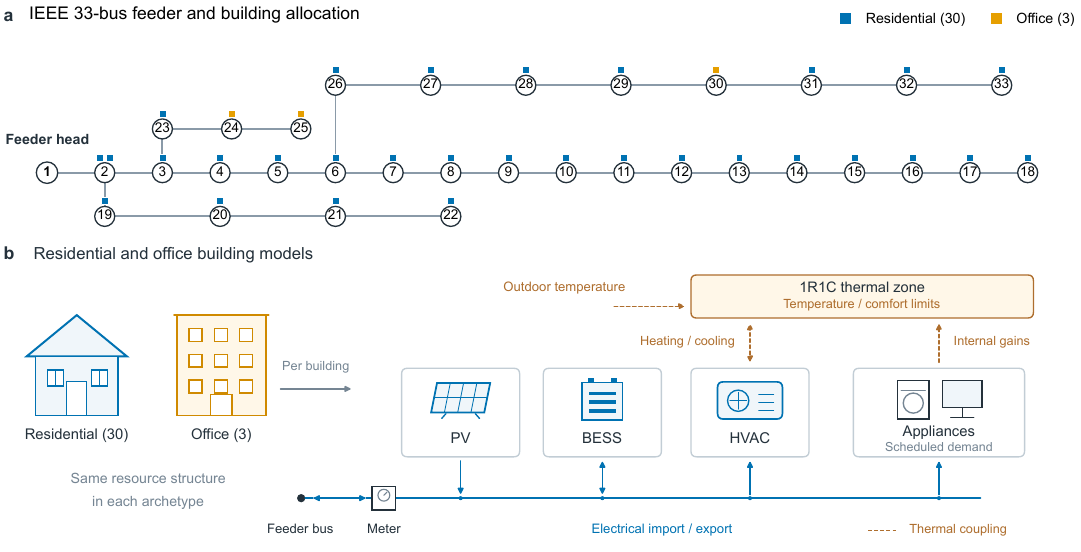}
    \caption{System configuration. (a) IEEE 33-bus feeder and building allocation. (b) Per-building resource model for both archetypes.}
    \label{fig:architecture}
    \vspace{-1.5em}
\end{figure*}

Forecast uncertainty further separates estimated flexibility potential from dependable operation.
Langner et al. compare deterministic, robust, and stochastic building controllers.
Their results demonstrate the importance of forecast errors for comfort and operating cost \cite{langner2024uncertainty}.
Zhang et al. coordinate mechanical voltage-control devices, reactive power, and active-power curtailment through stochastic control at multiple timescales \cite{zhang2021stochastic}.
Reliability-oriented DER coordination also shows why locally attractive operating decisions can conflict with distribution-system requirements \cite{navidi2023reliability}.
Nonlinear feasibility for a prescribed trajectory and reliability under uncertain disturbances therefore require distinct evidence.

Table~\ref{tab:positioning} summarizes how these approaches relate to the present study.
The paired lossless problems share a service equality.
The accepted feeder offer also compensates losses.
The reported retention is conditional on this proposal-and-acceptance construction.

\section{System and Service Model}

\subsection{Buildings and DERs}
The system contains $|\mathcal{B}|=33$ buildings on the 32 load buses of the balanced IEEE 33-bus radial feeder.
The population comprises 30 residential and three office archetypes scaled to 3.715~MW of aggregate design demand, as shown in Fig.~\ref{fig:architecture}(a).
The residential models do not represent individual dwellings.
Bus~2 hosts two residential buildings, and offices occupy buses~24, 25, and~30.
Each archetype includes scheduled appliance demand, ideal HVAC, PV, and a battery energy storage system (BESS), as illustrated in Fig.~\ref{fig:architecture}(b).
PV ratings total 2.972~MW; battery power and energy total 1.486~MW and 5.944~MWh.
A single-resistance, single-capacitance (1R1C) zone model represents the building thermal dynamics.
Positive power denotes import. Resource powers and energies use kW and kWh, with $R_b$ in K/kW, $C_b$ in kWh/K, and $\Delta t$ in hours.

For building $b$ and 15-min interval $k$, the discretized thermal model is
\begin{align}
 T_{b,k+1} &= a_b T_{b,k}+(1-a_b)T^{\mathrm{out}}_k
 +g_b(q^{\mathrm{int}}_{b,k}+q^{\mathrm{hvac}}_{b,k}),
 \label{eq:thermal}\\
 \underline T_{b,k}&\le T_{b,k+1}\le\overline T_{b,k},
 \qquad a_b=\exp[-\Delta t/(R_bC_b)],
 \label{eq:comfort}
\end{align}
where $R_b$ and $C_b$ denote lumped thermal resistance and capacitance, respectively, and $g_b=R_b(1-a_b)$.
The internal gain is $q^{\mathrm{int}}_{b,k}=0.7p^{\mathrm{app}}_{b,k}$, with no separate solar heat-gain term in this instance.
HVAC power satisfies $0\le p^{\mathrm{hvac}}_{b,k}\le\overline P_b^{\mathrm{hvac}}$.
The signed thermal input is $q^{\mathrm{hvac}}=3.5p^{\mathrm{hvac}}$ in heating mode and $q^{\mathrm{hvac}}=-3.0p^{\mathrm{hvac}}$ in cooling mode.
A nonzero reference action fixes the operating mode throughout the corresponding interval.

Battery charge $p^{\mathrm{ch}}$, discharge $p^{\mathrm{dis}}$, and energy $E$ satisfy
\begin{align}
 E_{b,k+1}&=E_{b,k}+\Delta t\left(\eta_b^{\mathrm{ch}}p^{\mathrm{ch}}_{b,k}
 -p^{\mathrm{dis}}_{b,k}/\eta_b^{\mathrm{dis}}\right), \label{eq:soc}\\
 \underline E_b&\le E_{b,k}\le\overline E_b,\quad
 0\le p^{\mathrm{ch}}_{b,k}\le z_{b,k}\overline P_b^{\mathrm{ch}},\nonumber\\
 0&\le p^{\mathrm{dis}}_{b,k}\le(1-z_{b,k})\overline P_b^{\mathrm{dis}},
 \quad z_{b,k}\in\{0,1\}. \label{eq:batteryexclusive}
\end{align}
The binary variable $z_{b,k}$ explicitly excludes simultaneous battery charging and discharging.
During activation, a battery keeps its reference charging or discharging mode and adjusts only its power level according to the service direction. An idle battery discharges for upward service and charges for downward service. These rules prevent charge–discharge cycling losses from being counted as flexibility. During recovery, binary constraints are used when needed to prevent simultaneous charging and discharging.
PV dispatch is limited by the available generation, with $0\le p^{\mathrm{pv}}_{b,k}\le p^{\mathrm{pv,av}}_{b,k}$.
The resulting building net import combines appliance demand with the controllable resource powers:
\begin{equation}
 p_{b,k}=p^{\mathrm{app}}_{b,k}+p^{\mathrm{hvac}}_{b,k}
 +p^{\mathrm{ch}}_{b,k}-p^{\mathrm{dis}}_{b,k}-p^{\mathrm{pv}}_{b,k}.
 \label{eq:buildingpower}
\end{equation}
Appliance and HVAC loads retain a power factor of 0.95, whereas PV and BESS operate at unity power factor.
The service assessment therefore includes active-power flexibility and excludes reactive-power control.

\subsection{Reference trajectory and tariff}
The reference input covers 12 months.
Sequential 24-h cost optimizations construct the reference.
Inputs include 15-min weather, deterministic archetype schedules, irradiance-scaled PV availability, and dynamic tariff profile.
The optimization interval is 15~min, with thermal and battery states continuous across daily boundaries.
This approach links daily optima rather than solving a single optimization over the complete input period.
At each site meter, imported energy is charged at $\pi_k$, and exported energy receives a credit of $\alpha\pi_k$:
\begin{equation}
 J^{\mathrm{energy}}=\sum_{b,k}\Delta t\left(\pi_k[p_{b,k}]_+
 -\alpha\pi_k[-p_{b,k}]_+\right).
 \label{eq:tariff}
\end{equation}
The base case uses an export-price factor of $\alpha=0.5$ throughout the evaluation.
Daily BESS energy neutrality prevents the reference optimization from borrowing stored energy across accounting boundaries.
The reference satisfies the base feeder envelope and passes nonlinear replay before flexibility is evaluated.
This zero-service check prevents baseline infeasibility from being attributed to flexibility erosion.

\subsection{Service contract and intertemporal closure}
For an hourly activation anchor $\tau$, direction $s\in\{+1,-1\}$, and duration $D\in\{15,60,240\}$ min, the initial lossless optimization proposes a constant amplitude $F\ge0$ through
\begin{equation}
 P^{0}_{0,k}-P_{0,k}=sF,
 \quad k\in\mathcal{K}(\tau,D),
 \label{eq:service}
\end{equation}
where $P_{0,k}=\sum_b p_{b,k}$ is lossless aggregate import in kW, and superscript~0 denotes the reference.
Nonlinear acceptance uses metered source import, including losses, and replaces this aggregate setpoint with a compensated one, as defined in Section~IV-C.
The sign convention assigns $s=+1$ to upward service and $s=-1$ to downward service.
Each optimization spans an 8-h horizon that includes activation and recovery.
At the end of the horizon, every zone temperature and BESS energy must return to its reference state within the declared tolerances:
\begin{equation}
 |T_{b,K}-T^0_{b,K}|\le\epsilon_T,\qquad
 |E_{b,K}-E^0_{b,K}|\le\epsilon_{E,b}.
 \label{eq:terminal}
\end{equation}
With $\mathrm{SOC}_b=E_b/E_b^{\mathrm{rated}}$, the external energy tolerance is $\epsilon_{E,b}=10^{-4}E_b^{\mathrm{rated}}$.
These terminal conditions include recovery in the service assessment and prevent unresolved state deviations from being shifted beyond the horizon.

\section{Feeder-Conditioned Flexibility Method}

\subsection{LinDistFlow network constraints}
For radial branch $(i,j)\in\mathcal{L}$, let $\mathcal{C}(j)$ denote the children of bus $j$ and let $p_{j,k},q_{j,k}$ denote its aggregate demand.
The lossless DistFlow approximation \cite{baran1989distflow} uses per-unit powers and impedances on a 10-MVA, 12.66-kV base:
\begin{align}
 P_{ij,k}&=p_{j,k}+\sum_{\ell\in\mathcal{C}(j)}P_{j\ell,k},\nonumber\\
 Q_{ij,k}&=q_{j,k}+\sum_{\ell\in\mathcal{C}(j)}Q_{j\ell,k},\label{eq:lindistflow}\\
 v_{j,k}&=v_{i,k}-2(r_{ij}P_{ij,k}+x_{ij}Q_{ij,k}).\nonumber
\end{align}
Here, $v$ is squared voltage magnitude, and the slack voltage is one p.u.
The network restrictions are
\begin{align}
 (0.95+\Delta V_{\mathrm{guard}})^2&\le v_{j,k}\le1.05^2,\label{eq:voltage}\\
 (P_{ij,k},Q_{ij,k})&\in\mathcal{S}_{ij},\label{eq:branch}\\
 (P_{01,k},Q_{01,k})&\in\mathcal{S}_{0},\quad
 S_{\mathrm{base}}P_{01,k}\ge-\overline P^{\mathrm{rev}}_0.\label{eq:head}
\end{align}
Each polygon imposes $P\cos\theta_m+Q\sin\theta_m\le\overline S\cos(\pi/16)/S_{\mathrm{base}}$, with $\theta_m=2\pi m/16$ and $m=0,\ldots,15$.
For branches, $\overline S_{ij}=\sqrt{3}(0.95)V_{\mathrm{base}}\overline I_{ij}$; for the transformer, $\overline S_0$ is its apparent-power rating.
The transformer restriction therefore includes reactive loading, rather than only an active-power ceiling.

The voltage guard is $\Delta V_{\mathrm{guard}}=0.0016$~p.u.
This value is rounded upward from twice the maximum LinDistFlow voltage overestimate in the complete yearly unmanaged nonlinear replay plus $10^{-4}$~p.u.
Branch, transformer, and reverse-flow ratings equal 120\% of their respective unmanaged peaks.
The resulting transformer and reverse-flow limits are 2890.945~kVA and 2687.978~kW.
These settings define a conservative assessment scenario; they are not an outer relaxation of the nonlinear feeder.

\subsection{Paired capacity problems and the bounded quantity}
Let $\mathcal{B}_c$ collect the building, resource, initial-state, mode, and recovery constraints for $(\tau,D,s)$, excluding the service equality.
Let $\mathcal{S}^{0}(F)$ impose \eqref{eq:service} with zero offset and let $\mathcal{L}_c$ impose \eqref{eq:lindistflow}--\eqref{eq:head}.
Suppressing contract indices, the paired lossless capacity problems are
\begin{align}
 \Fd&=\max_{F,\boldsymbol{x}}\{F:\boldsymbol{x}\in\mathcal{B}_c\cap\mathcal{S}^{0}(F)\},\label{eq:fd}\\
 \Fg_{\mathrm{lin}}&=\max_{F,\boldsymbol{x}}\{F:\boldsymbol{x}\in\mathcal{B}_c\cap\mathcal{S}^{0}(F)\cap\mathcal{L}_c\}.
 \label{eq:fg}
\end{align}
Both use identical references, resources, and recovery conditions.
Only in this lossless comparison are the network constraints the sole difference.

Stage~1 maximizes $F$.
Subsequent stages preserve $F\ge F^\star-\epsilon_F$, where $\epsilon_F=\max(0.1\,{\rm kW},0.001F^\star)$.
They minimize, in order,
\begin{equation}
 J_T\prec J^{\mathrm{energy}}\prec J_B\prec J_{\mathrm{PV}}\prec J_L,
 \label{eq:secondary}
\end{equation}
representing reference-temperature deviation, cost, battery throughput, PV curtailment, and a linear resistive-flow proxy.
Objective formulas, locks, and fallback rules are omitted for brevity.
Let $A_0$ be the feasible amplitude selected by this sequence for the feeder problem.
Nonlinear correction searches only $0\le F\le A_0$ and compensates losses as described below.

To define the bounded target precisely, let $\mathcal{N}(F)$ require nonlinear physical limits and source-side tracking of $F$ within the declared tolerances.
The capacity assessed after correction is
\begin{equation}
 \Fcap=\max_{F,\boldsymbol{x}}\{F:0\le F\le A_0,\quad
 \boldsymbol{x}\in\mathcal{B}_c\cap\mathcal{L}_c\cap\mathcal{N}(F)\}.
 \label{eq:fcap}
\end{equation}
The proposal budget $A_0$ is an explicit restriction of this target.
An accepted trajectory establishes $\Fglo\le\Fcap$.
By construction,
\begin{equation}
 \Fcap\le A_0\le\Fg_{\mathrm{lin}}\le\Fd,
 \qquad A_0\le\Fghi.
 \label{eq:boundchain}
\end{equation}
Here, $\Fghi$ retains the recorded primary ceiling, and $[\Fdlo,\Fdhi]$ brackets the demand-only optimum.
The search bracket's upper edge is never used as a capacity bound.
Equation~\eqref{eq:boundchain} follows from the imposed amplitude cap and the nested lossless problems.
It does not rely on nonlinear feasibility being contained within LinDistFlow.
Removing the cap or the conservative linear restrictions defines a different capacity, for which we claim no upper bound.

The conditional retention ratio $\rho=\Fcap/\Fd$ therefore satisfies
\begin{align}
 \rho^-&=\Fglo/\Fdhi,\nonumber\\
 \rho^+&=\min\{1,\Fghi/\Fdlo\},\nonumber\\
 e&=1-\rho\in[1-\rho^+,1-\rho^-].\label{eq:rho}
\end{align}
When $\Fdlo\le0.1$~kW but $\Fdhi>0.1$~kW, set $\rho^+=1$.
Ratios are undefined if $\Fdhi\le0.1$~kW.
These intervals account for proposal-budget conservatism, unresolved search gaps, and solver tolerances.
They do not quantify unavoidable loss of maximum nonlinear physical capacity.

\subsection{Loss compensation and nonlinear acceptance}
Every feeder candidate is replayed in the balanced nonlinear feeder functional mock-up unit (FMU), using its bus active and reactive powers.
Modelica and the Functional Mock-up Interface support reusable models and controller benchmarking \cite{wetter2013modelica,blum2021boptest,cupeiro2019cosimulation}.
Let $P^{\mathrm{AC}}_{0,k}=P_{0,k}+\ell_k$ denote source import including losses.
The metered directional response is $r^{\mathrm{AC}}_k=s(P^{\mathrm{AC},0}_{0,k}-P^{\mathrm{AC}}_{0,k})$.
For a lossless aggregate setpoint $s(P^0_{0,k}-P_{0,k})=F+\delta_k$,
\begin{equation}
 r^{\mathrm{AC}}_k=F+\delta_k+s(\ell^0_k-\ell_k),\qquad
 \delta_k^{j+1}=\delta_k^j+F-r^{\mathrm{AC},j}_k.
 \label{eq:lossfeedback}
\end{equation}
The bid amplitude is $F$, and acceptance requires $|r^{\mathrm{AC}}_k-F|\le\max(0.1\,{\rm kW},0.001F)$ during activation.
Starting from $\delta_k^0=0$, the implementation permits up to five fixed-amplitude redispatches when tracking is the only failed nonlinear test.
Any physical-limit, convergence, or optimization failure ends that fixed-amplitude attempt.
Demand-only replay is diagnostic and receives no loss compensation.
Its capacity therefore remains a lossless building budget rather than a separately optimized nonlinear source product.

If the initial amplitude cannot be accepted, the algorithm starts with the certified reference as a zero-amplitude witness and bisects $[0,A_0]$.
At each midpoint, it redispatches with \eqref{eq:lossfeedback}.
A pass replaces the retained witness and lower search endpoint; a failure reduces the upper search endpoint.
It stops at a tolerance of $\max(0.1\,{\rm kW},0.001A_0)$ or after 32 midpoint iterations.
The retained response is feasible.
Failure of one allocation does not establish that its amplitude is infeasible or that the result is globally maximal.
For $A_0=0$, this route returns the reference without midpoint bisection.

Thermal dynamics, battery SOC, comfort, power balance, and terminal states are independently recomputed from the exported schedule.
The optimization uses kWh-scaled battery balance and reserves a $10^{-6}$ SOC margin inside the $10^{-4}$ terminal tolerance.
SOC is reconstructed from charging and discharging powers.

\section{Experimental Design and Validation}

The experiment compares a lossless building budget with source offers accepted through nonlinear replay under the linear proposal budget in Section~IV.
Matched references and recovery conditions control for resource differences.
The comparison remains conditional on the conservative network formulation and the loss-compensation search.

\subsection{Experimental design}
We evaluated the 33-building population on the balanced IEEE 33-bus feeder described in Section~III.
For each service contract, demand-only and feeder-aware optimizations used the same initial states, cost-optimal reference, and available HVAC, PV, and BESS resources.
They also shared an 8-h activation--recovery horizon.
The feeder-aware problem added the network constraints and nonlinear certification procedure in Section~IV.
The reference was frozen only after complete interval coverage and successful building and nonlinear zero-service checks.
This prevented reference infeasibility from being counted as service-induced erosion.

The evaluation covered April--September 2018 at 15-min resolution, comprising 17,568 intervals.
Hourly activation anchors were included whenever the complete 8-h horizon fell within this period, yielding 4,385 anchors.
Combining every anchor with both directions and durations of 15, 60, and 240~min produced 26,310 paired contracts and 52,620 capacity rows.
Seasonal comparisons used April--May (spring), June--August (summer), and September (autumn).\footnote{Due to the limitation of data availability, the autumn subset therefore represents September alone.}
Table~\ref{tab:contract} specifies the common configuration, including 120\% feeder headroom and an export-price factor of 0.5.
Winter conditions, resource ablations, and alternative headroom or tariff settings were outside this evaluation.

\begin{table}[t]
\caption{Configuration and boundary conditions for the completed six-month simulation experiment.}
\label{tab:contract}
\centering
\footnotesize
\begin{tabular}{@{}>{\raggedright\arraybackslash}p{0.31\columnwidth}>{\raggedright\arraybackslash}p{0.64\columnwidth}@{}}
\toprule
Item & Base definition \\
\midrule
Time & Apr.--Sep. 2018; 15-min steps; 4,385 complete 8-h hourly anchors \\
Service & Lossless proposal; loss-compensated source acceptance; up/down; 15, 60, and 240 min \\
Reference/state & Sequential 24-h grid-admissible cost optima; continuous states; nonlinear zero-service pass; daily SOC neutrality; terminal $T$ and SOC return \\
Tariff & JEPX price; export factor 0.5 \\
Voltage & LinDistFlow floor 0.9516 p.u.; nonlinear certificate $0.95\le V_i\le1.05$ p.u. \\
Loading & 120\% of nonlinear unmanaged branch/transformer/reverse-flow peaks \\
DER power factor & Loads fixed power factor; PV/BESS unity power factor \\
Uncertainty & Deterministic perfect-information capacity assessment \\
\bottomrule
\end{tabular}
\vspace{-2em}
\end{table}

\subsection{Evaluation metrics and validation criteria}
Four groups of indicators connect the experimental objectives to the results.

\textbf{Capacity and deliverability}:
The intervals $[\Fdlo,\Fdhi]$ and $[\Fglo,\Fghi]$ bracket the lossless building capacity and the restricted target $\Fcap$ in kW.
The deliverability bounds $\rho^{-}$ and $\rho^{+}$ and the erosion interval in \eqref{eq:rho} measure conditional retention and erosion.
Scalar summaries use the conservative ratio $\rho^{-}$.
Ratios were undefined when $\Fdhi\le0.1$~kW.
These contracts were retained in capacity summaries and counted separately.
Intervals separate when $\Fghi<\Fdlo-\max(0.1\,{\rm kW},0.001\Fdhi)$, using the retained primary ceiling rather than a replay-search endpoint.
The frozen descriptive criterion required separation in at least 5\% of the contract population.
This was not a statistical significance test.

\textbf{Physical feeder feasibility}:
Voltage extrema and margins, maximum branch and transformer loading, reverse flow, and binding-constraint records identify the network limits encountered by each response.
For each physical test, the rejection fraction counts contracts with at least one violation anywhere in the 8-h trajectory; different mechanisms can overlap.
Demand-only and feeder-aware trajectories were checked against the same physical limits.
The nonlinear voltage band was 0.95--1.05~p.u., with acceptance tolerances of $10^{-4}$~p.u. for voltage and 0.1 percentage point for loading.

\textbf{Service delivery and building recovery}:
Maximum feeder-head tracking error measures the deviation of the nonlinear directional response from the requested constant amplitude during activation.
Acceptance used the larger of 0.1~kW and 0.1\% of that amplitude.
Capacity comparisons and certification used the same absolute and relative tolerance rule.
Service energy and signed rebound energy integrate the nonlinear directional response over activation and recovery, respectively.
Comfort and SOC checks assess internal feasibility.
Terminal deviations from reference states verify recovery within $10^{-3}$~$^{\circ}$C for temperature and $10^{-4}$ for SOC.
BESS throughput, PV curtailment, loss energy, and operating cost describe the resource and energy consequences of service.

\textbf{Certification and computational effort}:
The certification rate is the fraction of feeder-aware contracts passing all nonlinear, service, and building acceptance checks.
The archived recovery-route flag and the reduction from $A_0$ to $\Fglo$ quantify correction effort.
This flag includes zero-amplitude reference fallback; relative reductions are undefined for $A_0=0$.
Contract wall time, solver optimality gaps, and fallback or retry frequency measure computational effort and numerical reliability.
Infeasible, timed-out, and fallback runs were retained in the relevant denominators.

\section{Experimental results}

\begin{figure}[!t]
    \centering
    \includegraphics[width=\linewidth]{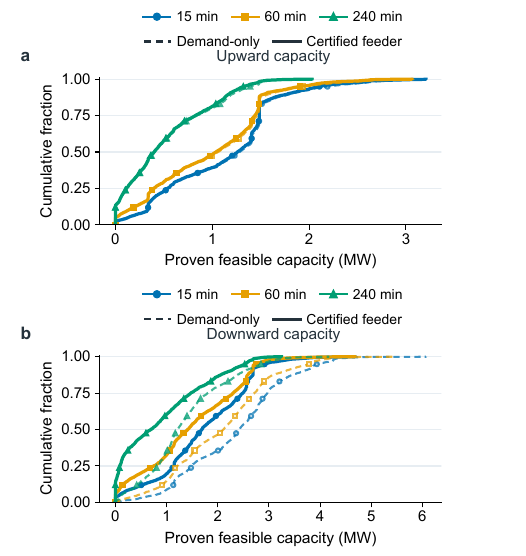}
    \caption{Accepted capacity lower endpoints across April--September 2018 for (a) upward and (b) downward service.}
    \label{fig:sixmonth}
\end{figure}

\begin{table}[t]
    \caption{Six-month median capacities and deliverability for each direction and service duration.
    ``Sep.'' denotes the percentage of contracts with separated capacity intervals.}
    \label{tab:sixmonth}
    \centering
    \scriptsize
    \setlength{\tabcolsep}{3.0pt}
    \begin{tabular}{@{}ccrrrr@{}}
    \toprule
    $s$ & $D$ & $\Fdlo$ & $\Fglo$ & $\rho^{-}$ & Sep. \\
     & (min) & (kW) & (kW) & & (\%) \\
    \midrule
    up & 15 & 1284 & 1261 & 0.999 & 4.36 \\
    up & 60 & 1061 & 1037 & 0.998 & 4.97 \\
    up & 240 & 406 & 400 & 0.998 & 6.41 \\
    down & 15 & 2421 & 1687 & 0.818 & 76.35 \\
    down & 60 & 2109 & 1413 & 0.812 & 73.82 \\
    down & 240 & 1218 & 695 & 0.721 & 71.06 \\
    \bottomrule
    \end{tabular}
    \vspace{-2em}
\end{table}

The simulation experiment completed all 26,310 service contracts and produced 52,620 paired capacity rows.
All 26,310 feeder-admissible trajectories passed the nonlinear FMU checks within the declared tolerances.

\subsection{Temporal and directional deliverability}
\begin{figure}[!t]
\centering
\includegraphics[width=\linewidth]{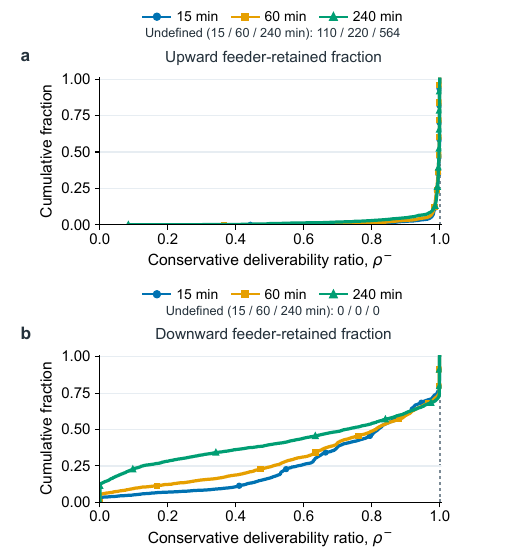}
\caption{Conservative feeder-retained fraction for (a) upward and (b) downward service.}
\label{fig:retained}
\end{figure}

Capacity retention differed markedly between upward and downward service (Figs.~\ref{fig:sixmonth} and~\ref{fig:retained}, and Table~\ref{tab:sixmonth}).
Median $\rho^{-}$ was 0.9989, 0.9980, and 0.9980 for upward service lasting 15, 60, and 240~min, respectively.
The corresponding downward medians were 0.8181, 0.8115, and 0.7208, with lower retention at longer durations.
For downward 240-min service, the interquartile range spanned 0.1261--0.9987, encompassing both severe erosion and near-complete retention.
All 894 undefined ratios arose from near-zero upward demand-only capacities.
The counts were 110, 220, and 564 in increasing order of duration.
These contracts remained in capacity summaries and were excluded only from ratio summaries.

Retaining the primary ceilings independently of bisection gave 10,391 separated intervals (39.49\%), exceeding the frozen 5\% descriptive criterion.
The upward fractions were 4.36\%, 4.97\%, and 6.41\%, compared with 76.35\%, 73.82\%, and 71.06\% downward, in increasing order of duration.
These comparisons concern \eqref{eq:fcap}; they do not identify unavoidable nonlinear physical capacity loss.

\begin{figure}[!t]
\centering
\includegraphics[width=\linewidth]{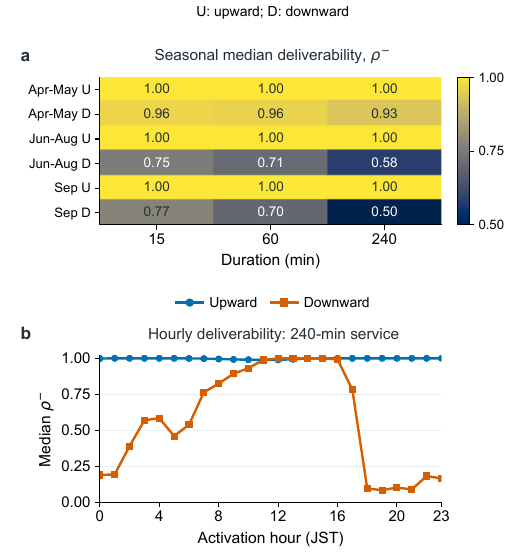}
\caption{Temporal variation in conservative deliverability. (a) Seasonal median ratios. (b) Hourly medians compare both directions of 240-min service.}
\label{fig:temporal}
\end{figure}

Seasonal and hourly summaries revealed further variation in downward 240-min service (Fig.~\ref{fig:temporal}).
Its median $\rho^{-}$ decreased from 0.931 in April--May to 0.581 in June--August and 0.501 in September.
The corresponding upward median remained 0.998 in each of these seasonal subsets.
Hourly downward medians were approximately one at 12:00--16:00 but no greater than 0.183 at 18:00--23:00.
Upward medians remained at least 0.987 across all hours of the day.

\subsection{Network exposure and active constraints}
\begin{figure}[!t]
\centering
\includegraphics[width=\linewidth]{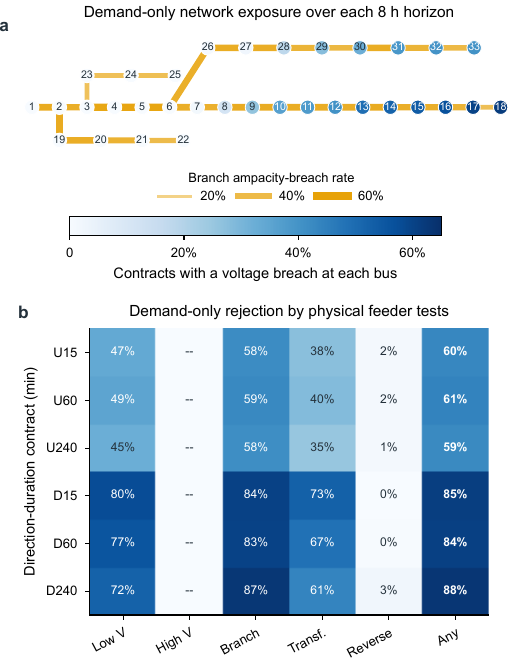}
\caption{Network exposure of demand-only trajectories across all contracts. (a) Bus color represents the fraction of 8-h trajectories breaching the local voltage band; branch width represents the ampacity-breach fraction. (b) Rejection fractions are shown by physical feeder test for upward (U) and downward (D) service.}
\label{fig:spatial}
\end{figure}

Figure~\ref{fig:spatial} reports nonlinear network exposure because aggregate optimization does not identify bus-level flexibility contributions.
At least one physical feeder test rejected 58.8--61.0\% of upward and 84.1--87.8\% of downward demand-only trajectories across the three durations.
Across all service contracts, the demand-only physical rejection rate was 72.7\%.
For instance, Bus~18 breached the voltage band in 61.7\% of demand-only contracts, and branch~4 exceeded its ampacity tolerance in 57.0\% of contracts.
Low voltage, branch ampacity, and transformer capacity were the dominant rejection mechanisms.
Mechanism frequencies overlapped, and no demand-only trajectory breached the upper voltage limit.

\subsection{Nonlinear delivery and recovery}
\begin{figure}[!t]
\centering
\includegraphics[width=\linewidth]{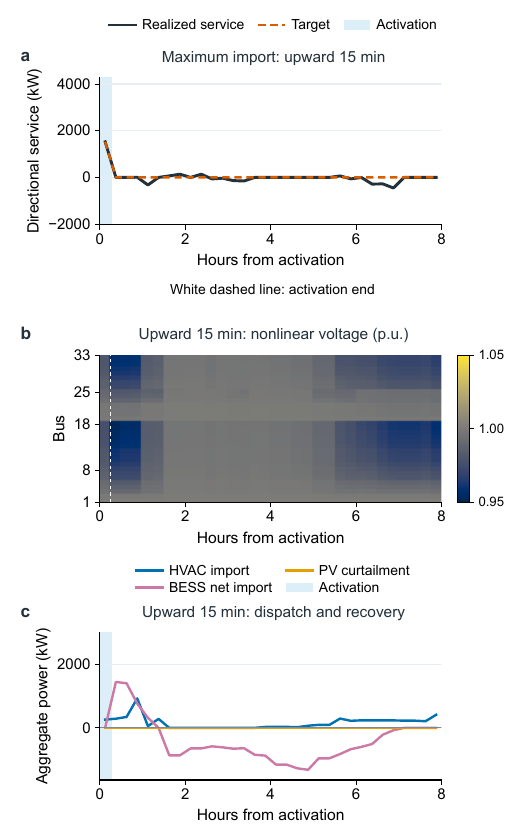}
\caption{Upward 15-min service at the preselected spring maximum-import anchor. (a) Delivered service and target, retaining signed post-event rebound. (b) Nonlinear voltage at all 33 buses on a 0.95--1.05~p.u. scale. (c) Aggregate HVAC electric power, BESS net import (negative during discharge), and PV curtailment over the full 8-h horizon.}
\label{fig:activation}
\end{figure}

\begin{figure}[!t]
\centering
\includegraphics[width=\linewidth]{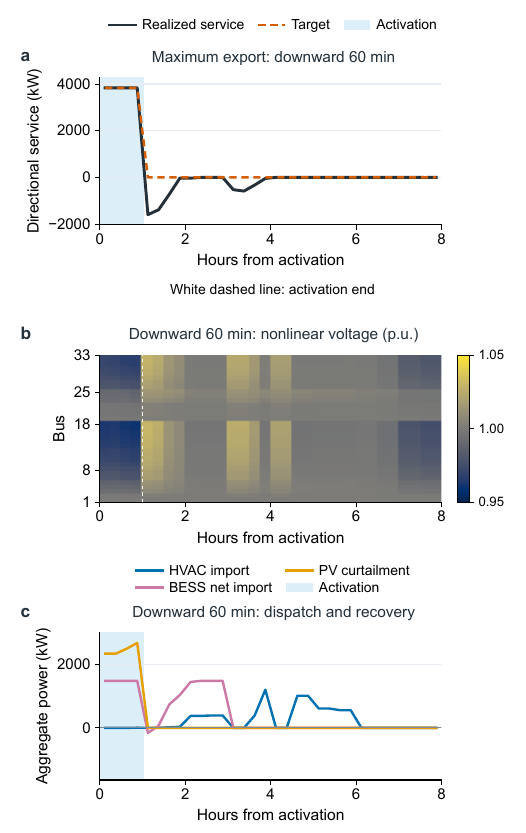}
\caption{Downward 60-min service at the preselected spring maximum-export anchor. (a) Delivered service and target, retaining signed post-event rebound. (b) Nonlinear voltage at all 33 buses. (c) Aggregate HVAC electric power, BESS net import (negative during discharge), and PV curtailment.}
\label{fig:activation-down}
\end{figure}

Figures~\ref{fig:activation} and~\ref{fig:activation-down} show the spring import and export extrema among the six preselected illustrative contracts.

At the maximum-import reference anchor, upward 15-min service delivered 1.526~MW with a maximum tracking error of 1.476~kW.
Voltage remained within 0.9527--1.0000~p.u., and maximum branch loading reached 99.0\% during the evaluated trajectory.
At the maximum-export anchor, downward 60-min service delivered 3.837~MW with a maximum tracking error of 0.615~kW.
Its voltage range was 0.9605--1.0293~p.u., and maximum branch loading reached 92.2\%.
Signed post-event rebound energies were $-346$ and $-1314$~kWh for the upward and downward cases, respectively.
Both trajectories satisfied terminal-state and nonlinear acceptance checks without capacity bisection.

\subsection{Certification and computation}
\begin{figure}[!t]
\centering
\includegraphics[width=\linewidth]{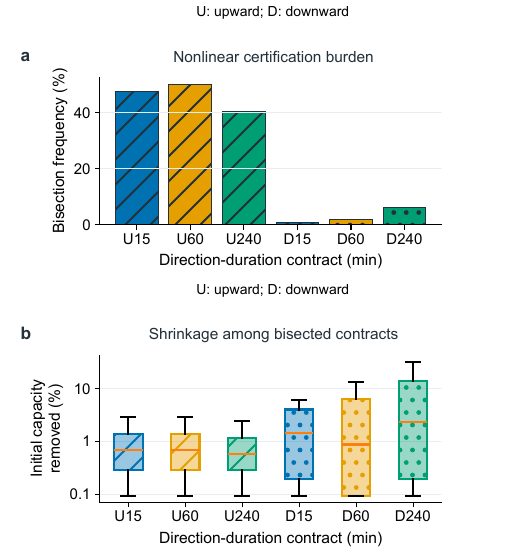}
\caption{Nonlinear certification across 26,310 service contracts. (a) Recovery-route frequency. (b) Relative capacity reduction for 6,443 recovery cases with positive initial amplitude (logarithmic scale above 0.1\%).}
\label{fig:robustness}
\end{figure}

At the first nonlinear replay, 2,127 proposals passed, and 24,183 failed source tracking alone.
None of these first rejections involved a physical feeder-limit test.
After loss compensation, 17,717 of the failed proposals were accepted at the original amplitude.
In total, 19,844 offers were accepted without amplitude reduction.
The other 6,466 contracts (24.58\%) entered the recovery route recorded by the archived bisection flag.
These included 23 zero-amplitude reference fallbacks, for which relative capacity reduction is undefined.
Thus, a high correction frequency can reflect source metering and redispatch even when the proposal budget is largely retained.
The upward recovery-route frequency was 40.6--50.0\%, compared with 0.87--6.34\% downward (Fig.~\ref{fig:robustness}(a)).
Among 6,443 recovery cases with positive initial amplitude, the median reduction was 0.684\%, the 95th percentile was 3.32\%, and the maximum was 97.27\% (Fig.~\ref{fig:robustness}(b)).

\begin{figure}[!t]
\centering
\includegraphics[width=\linewidth]{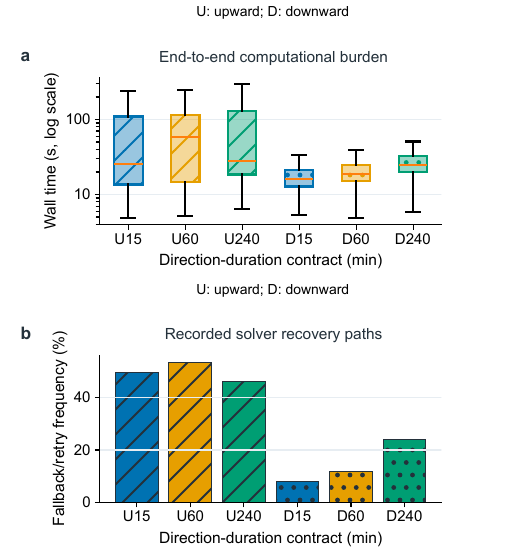}
\caption{Computation across 26,310 service contracts. (a) Contract wall time (logarithmic scale). (b) Solver fallback/retry frequency.}
\label{fig:computation}
\end{figure}

Median contract wall time was 21.6~s, and the 95th percentile was 155.0~s (Fig.~\ref{fig:computation}).
The longest contract required 3064.6~s because individual contracts could invoke multiple solver and certification attempts.
Solver fallback or retry paths occurred in 8,479 feeder contracts (32.2\%).
These were recovery paths rather than contract failures; all contracts remained in the reported denominators.
No contract ended in timeout.
The recorded primary-solve results only verify the optimization used to generate the initial proposal. They do not prove that the final nonlinear accepted offer is the globally maximum feasible service. In addition, all stored demand-only and feeder-admissible trajectories were independently checked to ensure that the building dynamics, comfort limits, battery constraints, and terminal recovery requirements were satisfied.

\section{Discussion and Conclusion}

This study assessed the conversion of building-side flexibility into deliverable feeder-head services.
In this study, we proposed a \textbf{paired building-only and feeder-aware flexibility assessment framework} to quantify how much building-side flexibility can be translated into deliverable feeder-head services under identical reference, resource, and recovery conditions.
The framework combines lossless capacity optimization with conservative feeder constraints, nonlinear Modelica replay, loss compensation, redispatch, and, when necessary, amplitude reduction.

The experimental results demonstrated that feeder constraints can substantially reduce deliverable flexibility, with a pronounced directional and temporal dependence: 
upward flexibility was almost fully retained, with median conservative retention close to 0.998 across all service durations, whereas downward flexibility was more strongly constrained and decreased to a median retention of 0.721 for 240-min services. 
The results also showed that many initial nonlinear replay failures were caused by feeder-loss-induced tracking errors rather than physical network violations, and that most of these cases could be corrected through redispatch without reducing the original service amplitude. 
Overall, the study shows that building-feasible flexibility and feeder-deliverable flexibility are not equivalent, and that reliable service assessment requires both network-aware capacity evaluation and nonlinear source-side validation.

\raggedcolsend
\raggedend
\bibliographystyle{IEEEtran}
\bibliography{references/references,references/context_references}

@article{oshaughnessy2022demand,
  author  = { O'Shaughnessy, Eric and Shah, Monisha and Parra, David and Ardani, Kristen },
  title   = {{The demand-side resource opportunity for deep grid decarbonization}},
  journal = {Joule},
  year    = {2022},
  volume  = {6},
  number  = {5},
  pages   = {972--983},
  doi     = {10.1016/j.joule.2022.04.010},
}

@article{navidi2023reliability,
  author  = { Navidi, Thomas and El Gamal, Abbas and Rajagopal, Ram },
  title   = {{Coordinating distributed energy resources for reliability can significantly reduce future distribution grid upgrades and peak load}},
  journal = {Joule},
  year    = {2023},
  volume  = {7},
  number  = {8},
  pages   = {1769--1792},
  doi     = {10.1016/j.joule.2023.06.015},
}

@article{baran1989distflow,
  author  = { Baran, M.E. and Wu, F.F. },
  title   = {{Network reconfiguration in distribution systems for loss reduction and load balancing}},
  journal = {IEEE Transactions on Power Delivery},
  year    = {1989},
  volume  = {4},
  number  = {2},
  pages   = {1401--1407},
  doi     = {10.1109/61.25627},
}

@article{fontenot2021b2dn,
  author  = { Fontenot, Hannah and Ayyagari, Krishna Sandeep and Dong, Bing and Gatsis, Nikolaos and Taha, Ahmad },
  title   = {{Buildings-to-distribution-network integration for coordinated voltage regulation and building energy management via distributed resource flexibility}},
  journal = {Sustainable Cities and Society},
  year    = {2021},
  volume  = {69},
  pages   = {102832},
  doi     = {10.1016/j.scs.2021.102832},
}

@article{mirakhorli2018mpc,
  author  = { Mirakhorli, Amin and Dong, Bing },
  title   = {{Model predictive control for building loads connected with a residential distribution grid}},
  journal = {Applied Energy},
  year    = {2018},
  volume  = {230},
  pages   = {627--642},
  doi     = {10.1016/j.apenergy.2018.08.051},
}

@article{jiang2018adn,
  author  = { Jiang, Tao and Li, Zening and Jin, Xiaolong and Chen, Houhe and Li, Xue and Mu, Yunfei },
  title   = {{Flexible operation of active distribution network using integrated smart buildings with heating, ventilation and air-conditioning systems}},
  journal = {Applied Energy},
  year    = {2018},
  volume  = {226},
  pages   = {181--196},
  doi     = {10.1016/j.apenergy.2018.05.091},
}

@article{zhang2022transformer,
  author  = { Zhang, Yin and Qian, Tong and Tang, Wenhu },
  title   = {{Buildings-to-distribution-network integration considering power transformer loading capability and distribution network reconfiguration}},
  journal = {Energy},
  year    = {2022},
  volume  = {244},
  pages   = {123104},
  doi     = {10.1016/j.energy.2022.123104},
}

@article{bay2022dmpc,
  author  = { Bay, Christopher J. and Chintala, Rohit and Chinde, Venkatesh and King, Jennifer },
  title   = {{Distributed model predictive control for coordinated, grid-interactive buildings}},
  journal = {Applied Energy},
  year    = {2022},
  volume  = {312},
  pages   = {118612},
  doi     = {10.1016/j.apenergy.2022.118612},
}

@article{yu2024online,
  author  = { Yu, Liang and Yue, Dong and Tan, Yangyang and Zhang, Shuang and Yang, Yu and Dou, Chunxia and Kuzin, Victor },
  title   = {{Online Distributed Coordination Operation for Grid-Interactive Efficient Residential Buildings}},
  journal = {IEEE Transactions on Smart Grid},
  year    = {2024},
  volume  = {15},
  number  = {4},
  pages   = {3639--3652},
  doi     = {10.1109/tsg.2023.3339761},
}

@article{zhang2021stochastic,
  author  = { Zhang, Zhengfa and da Silva, Filipe Faria and Guo, Yifei and Bak, Claus Leth and Chen, Zhe },
  title   = {{Double-layer stochastic model predictive voltage control in active distribution networks with high penetration of renewables}},
  journal = {Applied Energy},
  year    = {2021},
  volume  = {302},
  pages   = {117530},
  doi     = {10.1016/j.apenergy.2021.117530},
}

@article{rousseau2026envelope,
  author  = { Rousseau, Julie and Tajoli, Carlo and Cai, Hanmin and Heer, Philipp and Orehounig, Kristina and Hug, Gabriela },
  title   = {{Trajectory-Independent Flexibility Envelopes of Energy-Constrained Systems With State-Dependent Losses}},
  journal = {IEEE Transactions on Smart Grid},
  year    = {2026},
  volume  = {17},
  number  = {2},
  pages   = {1167--1179},
  doi     = {10.1109/tsg.2025.3620988},
}

@article{kheiri2026deliverable,
  author  = { Kheiri, Sevda Zeinal and Mirzaei, Mohammad Amin and Parvania, Masood },
  title   = {{Hierarchical energy flexibility coordination of all-electric buildings for deliverable grid services}},
  journal = {International Journal of Electrical Power \& Energy Systems},
  year    = {2026},
  volume  = {174},
  pages   = {111476},
  doi     = {10.1016/j.ijepes.2025.111476},
}

@article{lankeshwara2025doe,
  author  = { Lankeshwara, Gayan and Sharma, Rahul and Alam, M.R. and Yan, Ruifeng and Saha, Tapan K. },
  title   = {{Development and validation of a dynamic operating envelopes-enabled demand response scheme in low-voltage distribution networks}},
  journal = {Applied Energy},
  year    = {2025},
  volume  = {382},
  pages   = {125150},
  doi     = {10.1016/j.apenergy.2024.125150},
}

@article{gerdroodbari2022pq,
  author  = { Zabihinia Gerdroodbari, Yasin and Khorasany, Mohsen and Razzaghi, Reza },
  title   = {{Dynamic PQ Operating Envelopes for prosumers in distribution networks}},
  journal = {Applied Energy},
  year    = {2022},
  volume  = {325},
  pages   = {119757},
  doi     = {10.1016/j.apenergy.2022.119757},
}

@article{liu2021operating,
  author  = { Liu, Michael Z. and Ochoa, Luis Nando and Riaz, Shariq and Mancarella, Pierluigi and Ting, Tian and San, Jack and Theunissen, John },
  title   = {{Grid and Market Services From the Edge: Using Operating Envelopes to Unlock Network-Aware Bottom-Up Flexibility}},
  journal = {IEEE Power and Energy Magazine},
  year    = {2021},
  volume  = {19},
  number  = {4},
  pages   = {52--62},
  doi     = {10.1109/mpe.2021.3072819},
}

@article{liu2022opf,
  author  = { Liu, Michael Z and Ochoa, Luis F. and Wong, Peter K. C. and Theunissen, John },
  title   = {{Using OPF-Based Operating Envelopes to Facilitate Residential DER Services}},
  journal = {IEEE Transactions on Smart Grid},
  year    = {2022},
  volume  = {13},
  number  = {6},
  pages   = {4494--4504},
  doi     = {10.1109/tsg.2022.3188927},
}

@article{jiang2025trading,
  author  = { Jiang, Zhisen and Guo, Ye and Wang, Jianxiao },
  title   = {{Dynamic operating envelopes embedded peer-to-peer-to-grid energy trading}},
  journal = {Applied Energy},
  year    = {2025},
  volume  = {377},
  pages   = {124554},
  doi     = {10.1016/j.apenergy.2024.124554},
}

@article{li2023kpi,
  author  = { Li, Han and Johra, Hicham and de Andrade Pereira, Flavia and Hong, Tianzhen and Le Dr{\'e}au, J{\'e}r{\^o}me and Maturo, Anthony and Wei, Mingjun and Liu, Yapan and Saberi-Derakhtenjani, Ali and Nagy, Zoltan and Marszal-Pomianowska, Anna and Finn, Donal and Miyata, Shohei and Kaspar, Kathryn and Nweye, Kingsley and O'Neill, Zheng and Pallonetto, Fabiano and Dong, Bing },
  title   = {{Data-driven key performance indicators and datasets for building energy flexibility: A review and perspectives}},
  journal = {Applied Energy},
  year    = {2023},
  volume  = {343},
  pages   = {121217},
  doi     = {10.1016/j.apenergy.2023.121217},
}

@article{li2021review,
  author  = { Li, Han and Wang, Zhe and Hong, Tianzhen and Piette, Mary Ann },
  title   = {{Energy flexibility of residential buildings: A systematic review of characterization and quantification methods and applications}},
  journal = {Advances in Applied Energy},
  year    = {2021},
  volume  = {3},
  pages   = {100054},
  doi     = {10.1016/j.adapen.2021.100054},
}

@article{kathirgamanathan2020indicators,
  author  = { Kathirgamanathan, Anjukan and P{\'e}an, Thibault and Zhang, Kun and De Rosa, Mattia and Salom, Jaume and Kummert, Micha{\"e}l and Finn, Donal P. },
  title   = {{Towards standardising market-independent indicators for quantifying energy flexibility in buildings}},
  journal = {Energy and Buildings},
  year    = {2020},
  volume  = {220},
  pages   = {110027},
  doi     = {10.1016/j.enbuild.2020.110027},
}

@article{tang2021flexibility,
  author  = { Tang, Hong and Wang, Shengwei and Li, Hangxin },
  title   = {{Flexibility categorization, sources, capabilities and technologies for energy-flexible and grid-responsive buildings: State-of-the-art and future perspective}},
  journal = {Energy},
  year    = {2021},
  volume  = {219},
  pages   = {119598},
  doi     = {10.1016/j.energy.2020.119598},
}

@article{bampoulas2021framework,
  author  = { Bampoulas, Adamantios and Saffari, Mohammad and Pallonetto, Fabiano and Mangina, Eleni and Finn, Donal P. },
  title   = {{A fundamental unified framework to quantify and characterise energy flexibility of residential buildings with multiple electrical and thermal energy systems}},
  journal = {Applied Energy},
  year    = {2021},
  volume  = {282},
  pages   = {116096},
  doi     = {10.1016/j.apenergy.2020.116096},
}

@article{stinner2016thermal,
  author  = { Stinner, Sebastian and Huchtemann, Kristian and M{\"u}ller, Dirk },
  title   = {{Quantifying the operational flexibility of building energy systems with thermal energy storages}},
  journal = {Applied Energy},
  year    = {2016},
  volume  = {181},
  pages   = {140--154},
  doi     = {10.1016/j.apenergy.2016.08.055},
}

@article{zhang2018consumer,
  author  = { Zhang, Yang and Campana, Pietro Elia and Yang, Ying and Stridh, Bengt and Lundblad, Anders and Yan, Jinyue },
  title   = {{Energy flexibility from the consumer: Integrating local electricity and heat supplies in a building}},
  journal = {Applied Energy},
  year    = {2018},
  volume  = {223},
  pages   = {430--442},
  doi     = {10.1016/j.apenergy.2018.04.041},
}

@article{hwang2023hosting,
  author  = { Hwang, Hyunkyeong and Yoon, Ahyun and Yoon, Yongtae and Moon, Seungil },
  title   = {{Demand response of HVAC systems for hosting capacity improvement in distribution networks: A comprehensive review and case study}},
  journal = {Renewable and Sustainable Energy Reviews},
  year    = {2023},
  volume  = {187},
  pages   = {113751},
  doi     = {10.1016/j.rser.2023.113751},
}

@article{langner2024uncertainty,
  author  = { Langner, Felix and Wang, Weimin and Frahm, Moritz and Hagenmeyer, Veit },
  title   = {{Model predictive control of distributed energy resources in residential buildings considering forecast uncertainties}},
  journal = {Energy and Buildings},
  year    = {2024},
  volume  = {303},
  pages   = {113753},
  doi     = {10.1016/j.enbuild.2023.113753},
}

@article{wetter2013modelica,
  author  = { Wetter, Michael and Zuo, Wangda and Nouidui, Thierry S. and Pang, Xiufeng },
  title   = {{Modelica Buildings library}},
  journal = {Journal of Building Performance Simulation},
  year    = {2013},
  volume  = {7},
  number  = {4},
  pages   = {253--270},
  doi     = {10.1080/19401493.2013.765506},
}

@article{blum2021boptest,
  author  = { Blum, David and Arroyo, Javier and Huang, Sen and Drgo{\v n}a, J{\'a}n and Jorissen, Filip and Walnum, Harald Taxt and Chen, Yan and Benne, Kyle and Vrabie, Draguna and Wetter, Michael and Helsen, Lieve },
  title   = {{Building optimization testing framework (BOPTEST) for simulation-based benchmarking of control strategies in buildings}},
  journal = {Journal of Building Performance Simulation},
  year    = {2021},
  volume  = {14},
  number  = {5},
  pages   = {586--610},
  doi     = {10.1080/19401493.2021.1986574},
}

@article{cupeiro2019cosimulation,
  author  = { Reinbold, Vincent and Protopapadaki, Christina and Tavella, Jean-Philippe and Saelens, Dirk },
  title   = {{Assessing scalability of a low-voltage distribution grid co-simulation through functional mock-up interface}},
  journal = {Journal of Building Performance Simulation},
  year    = {2019},
  volume  = {12},
  number  = {5},
  pages   = {637--649},
  doi     = {10.1080/19401493.2019.1597923},
}

\end{document}